\documentclass[conference,letterpaper]{IEEEtran}
\IEEEoverridecommandlockouts
\usepackage[utf8]{inputenc}
\usepackage[T1]{fontenc}
\usepackage{url}
\usepackage{ifthen}
\usepackage{cite}
\usepackage[cmex10]{amsmath}

\usepackage{cite}
\usepackage{amsmath,amssymb,amsfonts}

\usepackage{graphicx}
\usepackage{textcomp}
\usepackage{xcolor}
\usepackage{threeparttable}
\usepackage{bm}
\usepackage{amsfonts}
\usepackage{booktabs}
\usepackage{multirow}
\usepackage{subfigure}
\usepackage{tabularx}
\usepackage{algorithm}
\usepackage{amsmath}
\usepackage{amssymb}
\usepackage{color}
\usepackage{cite}

\usepackage{algpseudocode}
\usepackage{lineno,hyperref}
\usepackage{graphicx}
\usepackage{epsfig}
\usepackage{array}

\usepackage{graphicx}
\usepackage{epsfig}
\usepackage{graphicx}
\usepackage{amsmath}
\usepackage{amssymb}
\usepackage{multirow}
\usepackage{changepage}
\usepackage{subfigure}

\newcommand{\PreserveBackslash}[1]{\let\temp=\\#1\let\\=\temp}
\newcolumntype{C}[1]{>{\PreserveBackslash\centering}p{#1}}

\newtheorem{theorem}{Theorem}

\def\BibTeX{{\rm B\kern-.05em{\sc i\kern-.025em b}\kern-.08em
    T\kern-.1667em\lower.7ex\hbox{E}\kern-.125emX}}
\begin{document}
\title{Optimized Rate-Profiling for PAC Codes}

\author{\IEEEauthorblockN{He Sun}
\IEEEauthorblockA{\textit{School of Electronic and} \\
\textit{Information Engineering} \\
\textit{Beihang University}\\
Beijing, China \\
sunhe1710@buaa.edu.cn}
\and
\IEEEauthorblockN{Emanuele Viterbo}
\IEEEauthorblockA{\textit{Department of Electrical and Computer} \\
\textit{Systems Engineering (ECSE)} \\
\textit{Monash University}\\
Clayton, Melbourne, Australia\\
emanuele.viterbo@monash.edu}
\and
\IEEEauthorblockN{Rongke Liu}
\IEEEauthorblockA{\textit{School of Electronic and} \\
\textit{Information Engineering} \\
\textit{Beihang University}\\
Beijing, China\\
rongke\_liu@buaa.edu.cn}
}
\maketitle
\begin{abstract}The polarization-adjusted convolutional (PAC) codes concatenate the polar transform and the convolutional transform to improve the decoding performance of the finite-length polar codes, where the rate-profile is used to construct the PAC codes by setting the positions of frozen bits. However, the optimal rate-profile method of PAC codes is still unknown.

In this paper, an optimized rate-profile algorithm of PAC codes is proposed. First, we propose the \emph{normalized compression factor} (NCF) to quantify the transmission efficiency of useful information, showing that the distribution of useful information that needs to be transmitted after the convolutional transform should be adaptive to the capacity profile after finite-length polar transform. This phenomenon indicates that the PAC code improves the transmission efficiency of useful information, which leads to a better decoding performance than the polar codes with the same length. Then, we propose a novel rate-profile method of PAC codes, where a quadratic optimization model is established and the Euclidean norm of the NCF spectrum is adopted to construct the objective function. Finally, a heuristic bit-swapping strategy is designed to search for the frozen set with high objective function values, where the search space is limited by considering the only bits with medium Hamming weight of the row index. Simulation results show that the PAC codes with the proposed optimized rate-profile construction have better decoding performance than the PAC codes with the originally proposed Reed-Muller design construction.
\end{abstract}


\IEEEpeerreviewmaketitle

\section{Introduction}

\IEEEPARstart {P}{olar} codes are the first family of codes that provably achieve the capacity of a variety of channels with a simple encoder and decoder\cite{POLAR}. However, due to insufficient polarization and sub-optimal decoding algorithms\cite{ISC}, the finite-length decoding performance of polar codes is not ideal. In \cite{PAC}, the polarization-adjusted convolutional (PAC) codes are proposed to improve the decoding performance of finite-length polar codes, which are groundbreaking for the optimization of finite-length polar codes. With list decoding and sequential decoding\cite{PACRA1}, Viterbi decoding\cite{PACRA0} and successive-cancellation list (SCL)\cite{PACSCLP} decoding, the PAC codes show better decoding performance than polar codes with the same code length.

In PAC codes, the source codes are assigned into a longer sequence to obtain the data carrier sequence, where redundant zeros are padded among the source bits. The positions of the padded zeros are called the frozen set. The process of padding redundant bits to obtain the input sequence of PAC codes is called the \emph{rate-profile}. After rate-profile, the input sequence is encoded by convolutional transform and polar transform to obtain the codeword. Similar with polar codes, the selection of frozen bits is important for the construction of PAC codes. However, the optimal rate-profile criterion that can guide this selection is an open problem.

In this paper, we propose novel algorithms to analyze and optimize the rate-profile of PAC codes, where a quadratic optimization model is established for the optimal selection of frozen bits and a swapping algorithm is designed to select the optimal frozen sets. The main contributions are summarized as follows,
\begin{itemize}
  \item An analysis method of the convolution-polar transform is established, where the \emph{normalized compression factor} is proposed to quantify the transmission efficiency of useful information. With the proposed compression factor, we find that the distribution of useful information carried by the input sequence of PAC codes is adaptive to the distribution of channel capacity after the finite-length polarization, which leads to a better error-correction performance of PAC codes than the polar codes. Further, a large compression factor means a high transmission efficiency of useful information, which provides a heuristic basis for the optimization of PAC codes.
  \item An optimized rate-profile algorithm of PAC codes is proposed. Based on the Euclidean norm of the compression factors, the optimization of rate-profile is modeled as a quadratic optimization model. Then, the optimized rate-profile pattern is selected by a bit-swapping searching algorithm. In order to limit the search space, we propose to select some bits with medium Hamming weight of row index as the search space. Simulation results show that the decoding performance of PAC codes is improved with the optimized rate-profile algorithm.
\end{itemize}

The paper is organized as follows. Section \ref{sec2} introduces the background and the notations for PAC codes. Section \ref{sec3} analyzes the polarization-adjusted convolutional transform and proposes the compression factor. In Section \ref{sec4}, the optimized rate-profile of PAC codes is designed. Simulation results are given in Section \ref{sec5}. Section \ref{sec6} provides the conclusions.

\section{Preliminaries}\label{sec2}

\begin{figure}[ht]
\begin{center}
\centering
\includegraphics[scale=0.2195]{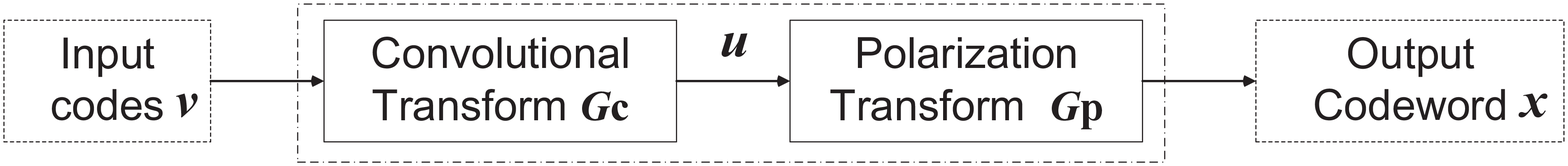}
\caption{The encoding procedure of the PAC codes.}
\label{F001}
\end{center}
\end{figure}

The encoding procedure of PAC codes consists of the convolutional encoding operation and the polar encoding operation, which is shown in Fig. \ref{F001}.

For a PAC code with generator polynomial $\textbf{g}=[g_0,g_1,\cdots,g_m]$, the convolutional transform can be represented in matrix form by
\begin{equation}
\textbf{u} = \textbf{v}{\textbf{G}_c},
\label{eq1}\end{equation}
where $\textbf{v}=[v_1,v_2,\cdots,v_N]$ is the input sequence of the convolutional encoder, $\textbf{u}=[u_1,u_2,\cdots,u_N]$ is the output sequence of the convolutional encoder, and $\textbf{G}_c$ is an upper-triangular Toeplitz\cite{ToeplitzM} matrix.
The polar transform can be denoted by
\begin{equation}
\textbf{x} = \textbf{u}{\textbf{G}_p},
\label{eq2}\end{equation}
where $\textbf{x}=[x_1,x_2,\cdots,x_N]$ is the output sequence of the polar encoder, and $\textbf{G}_p$ is the generator matrix of polar codes. A PAC code of length $N$ is generated by
\begin{equation}
\left. \begin{array}{l}
\textbf{u} = \textbf{v}{\textbf{G}_c}\\
\textbf{x} = \textbf{u}{\textbf{G}_p}
\end{array} \right\} \Leftrightarrow \textbf{x} = \textbf{v}\textbf{G},
\label{eq3}\end{equation}
where $\textbf{G} = {\textbf{G}_c}{\textbf{G}_p}$ denotes the generator matrix of the PAC codes. The input sequence $\textbf{v}$ with length $N$ is obtained by the rate-profile operation, which assigns the $K$ information bits and $N-K$ zeros into the vector $\textbf{v}$. The $N-K$ zeros are named as frozen bits, which are known by both the encoder and the decoder. The remaining $K$ non-frozen bits carry the useful messages that need to be transmitted to the receiver.

Similar with polar codes, the selection of frozen sets tremendously determines the decoding performance. For polar codes, different methods are designed to select the frozen sets\cite{POLAR}, \cite{E2015}, \cite{GAS}, \cite{PWS}, \cite{VARDYCONSTRUCTION}. For PAC codes, the rate-profile scheme\cite{PAC} based on the \emph{Reed-Muller (RM) design} adopts the Hamming weight of the row indexes to determine the frozen set of the PAC codes. Another rate-profile algorithm of PAC codes uses the capacity and the cut-off rates of polar codes\cite{ARIORIGIN} to select frozen sets, which is called the polar design. Although the PAC codes with RM design show better decoding performance than that with the polar design, the optimal rate-profile algorithm of the PAC codes remains unknown.

\section{Analysis of the Convolution-polar Transform}\label{sec3}

The PAC codes achieve a more flexible rate assignment, which is adaptable to the capacity distribution after the finite-length polar transform and improves the error-correction ability. In this section, we first provide a sight into the advantages of PAC codes in terms of the transmission efficiency of useful messages. In order to analyze the transmission efficiency of useful messages, we evaluate the proportion of non-frozen bits transmitted on each coded bit. During the convolutional transform, each coded bit is generated by a column of the generator matrix $\textbf{G}_c$, which can be represented by the equation
\begin{equation}
\begin{split}
{u_i} = {v_{{j_1}}} \oplus {v_{{j_2}}} \oplus  \cdots  \oplus {v_{{{j_p}_i}}}.
\end{split}\label{eq005}
\end{equation}

\begin{figure}[ht]
\begin{center}\hspace{-0.381cm}
\centering
\includegraphics[scale=0.3965]{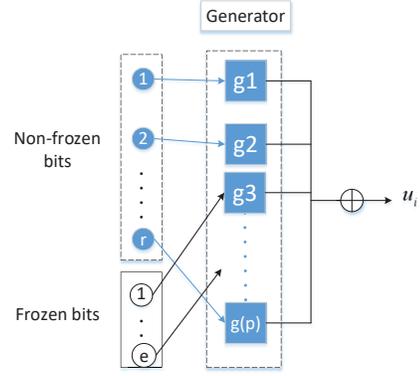}
\caption{Encoding of $u_i$ by the convolutional transform.}
\label{F002}
\end{center}
\end{figure}

The $\left\{ {{j_1} \cdots {{j_p}_i}} \right\}$ denotes the row index of the elements with value one in the $i$-th column of the generator matrix $\textbf{G}_c$, where $p_i$ is the Hamming weight of the $i$-th column of $\textbf{G}_c$. Eq. (\ref{eq005}) depicts that the source bits (${v_{{j_1}}}, {v_{{j_2}}},  \cdots, {v_{{{j_p}_i}}}$) are compressed and transmitted by the $i$-th channel corresponding to $u_i$ after the convolutional transform. For example, Fig. \ref{F002} shows the encoding procedure of $u_i$, where the operation is determined by the $i$-th column of $\textbf{G}_c$ whose coefficients are denoted by the blue rectangles. The source bits consist of $r_i$ non-frozen bits and $e_i$ frozen bits, where $e_i=p_i-r_i$.

After encoding, the source bits are compressed into $u_i$ and transmitted by a subchannel $W:{u_i} \mapsto x_1^N$, where the subchannel is generated by the polar transform. The number $r$ is determined by the frozen set of the PAC codes.

Since PAC codes have frozen bits, the number of non-frozen bits which are compressed into $u_i$ is upper bounded by ${r_i} {\leq} {p_i}$. For each bit $u_i$, the \emph{normalized compression factor} (NCF) is
\begin{equation}
\begin{split}
\gamma\left( {{u_i}} \right) \buildrel \Delta \over = \frac{{r_i}}{p_i},
\end{split}
\end{equation}
where $\gamma$ denotes the NCF. The NCF denotes the proportion of non-frozen bits compressed into bit $u_i$. The bit $u_i$ with a high compression factor carries a large proportion of the useful information, and each bit $v_i$ obtains a small proportion of the channel capacity. Therefore, a good channel is needed to transmit the bit $u_i$. For example, we consider a column with length $N=4$ and Hamming weight $p=2$. The generator equation of $u_i$ is
\begin{equation}
\begin{split}
\left[ {{v_1},{v_2},{v_3},{v_4}} \right]\left[ {\begin{array}{*{20}{c}}
1\\
1\\
0\\
0
\end{array}} \right] = {v_1} \oplus {v_2}.
\end{split}\label{eq0091}
\end{equation}

Three cases are considered as follows,
\begin{itemize}
  \item Case 1. When $v_1, v_2\in \cal F$, the number of compressed non-frozen bits is $r=0$ and the NCF is $\gamma=0$, which means that no useful information is transmitted by $u_i$. Hence, a channel with low capacity is needed for $u_i$.
  \item Case 2. When $v_1 \in \cal F$ and $v_2 \in {\cal F}^c$, the number of compressed non-frozen bits is $r=1$ and the NCF is $\gamma=0.5$, which means that the useful information of one non-frozen bit is transmitted by $u_i$. Therefore, a good channel should be assigned for $u_i$.
  \item Case 3. When $v_1, v_2 \in {\cal F}^c$, $r=2$ and the NCF is $\gamma=1$, which means that the useful information from two non-frozen bits is transmitted by $u_i$. Therefore, a better channel is needed for $u_i$ than that used in case 2.
\end{itemize}

\begin{figure}[ht]
\begin{center}\hspace{-0.381cm}
\centering
\includegraphics[scale=0.433695]{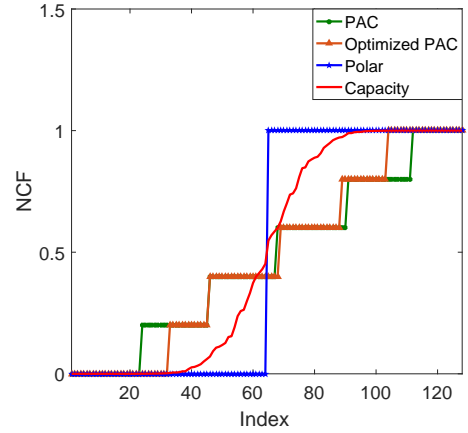}
\caption{NCF distribution of sequence $\textbf{u}$ input into the polar transform in PAC and polar codes ($N=128$). The capacity is obtained in binary erasure channel (BEC) with error probability $\varepsilon=0.5$. The curves are sorted results.}
\label{F002QI}
\end{center}
\end{figure}

Fig. \ref{F002QI} indicates the NCF distribution of $\textbf{u}$ in the PAC codes and polar codes, where the $\textbf{u}$ in PAC codes is obtained by the rate-profile and the convolutional transform. For polar codes, the input bit $u_i$ is either frozen bit or non-frozen bit, which means that the NCF of $u_i$ is either 0 or 1. However, for finite-length polar codes, the polarization is not ideal, and the capacity of the polarized subchannel is between 0 and 1, not strictly equal to 0 or 1. The frozen bits on the relatively unreliable subchannels carry no useful messages, which leads to a waste of the channel capacity. Meanwhile, since the capacity of the reliable subchannel with finite-length polarization is smaller than 1, bit errors might occur in the transmission of non-frozen bits.

Different from the polar codes, the PAC codes adopt a rate-profile and convolutional transform before the polar transform, which leads to a flexible NCF assignment for the input sequence of the polar transform. After the convolutional transform, there are a lot of bits in $\textbf{u}$ with medium NCF values. The distribution of the compressed information that we need to transmit in $\textbf{u}$ better matches the capacity distribution after the finite-length polar transform, and the utilization of channel capacity is improved, which leads to a better error-correction performance of PAC codes than the finite-length polar codes. In addition, for PAC codes, each non-frozen bit $v_i \in {\cal F}^c$ might be re-transmitted by multiple bits in the sequence $\textbf{u}$. Although some reliable polarized subchannels are not ideally noiseless channels after the finite-length polar transform, the multiple re-transmissions of $v_i$ improve the transmission reliability. For a generator polynomial $\textbf{g}$ with Hamming weight $w$, each non-frozen bit will be transmitted by $w$ bits of sequence $\textbf{u}$. The amount of useful information of non-frozen bit $v_i$ transmitted in a certain bit $u_j$ is determined by the NCF of $u_j$.

The NCF is a heuristic measure of the transmission efficiency of useful messages. For a given generator matrix and frozen set, the NCF corresponding to each output bit can be calculated, which is called the NCF spectrum. For the PAC codes, the number of non-frozen bits corresponding to each bit in $\textbf{x}$ can be obtained in matrix form by
\begin{equation}
\begin{split}
r \left( \textbf{x} \right) = r \left( \textbf{v} \right) \times \textbf{G},
\end{split}\label{eq008}
\end{equation}
where $r \left( \textbf{x} \right)$ denotes the vector $\left[ {r \left( {{x_1}} \right),r \left( {{x_2}} \right), \cdots ,r \left( {{x_N}} \right)} \right]$, and the notation $\times$ denotes the decimal product. $r(\textbf{v})$ is a vector that denotes the number of non-frozen bits in each element of vector $\textbf{v}$. It can be obtained by
\begin{equation}
\begin{split}
r\left( {{v_i}} \right) = \left\{ {\begin{array}{*{20}{c}}
1&{i \in {{\cal F}^c},}\\
0&{i \in {\cal F}.}
\end{array}} \right.
\end{split}\label{eq009}
\end{equation}

Then the Hamming weight of the columns of the generator matrix $\textbf{G}$ can be calculated by
\begin{equation}
\begin{split}
\textbf{p} = \left[ {\text{sum}\left[ {\textbf{G}\left( {:,1} \right)} \right],\text{sum}\left[ {\textbf{G}\left( {:,2} \right)} \right], \cdots ,\text{sum}\left[ {\textbf{G}\left( {:,N} \right)} \right]} \right].
\end{split}\label{eq0010}
\end{equation}

With the number of non-frozen bits in (\ref{eq008}) and the Hamming weight of each column obtained by (\ref{eq0010}), the NCF spectrum of $\textbf{x}$ can be obtained by
\begin{equation}
\begin{split}
\gamma\left( \textbf{x} \right) = r\left( \textbf{v} \right) \times {\tilde {\textbf{G}}},
\end{split}\label{eq00111}
\end{equation}
where the compression coefficient matrix ${\tilde {\textbf{G}}}$ is obtained by dividing the elements in the $i$-th column of matrix $\textbf{G}$ by $p_i$ for $i=1,2,\cdots,N$. The $\tilde{\textbf{G}}$ can be calculated by
\begin{equation}
\begin{split}
\tilde {\textbf{G}}{\rm{ = }}{\textbf{G}} \times \text{diag}\left( {\tilde{{\textbf{p}}}} \right),
\end{split}\label{eq00113}
\end{equation}
where ${\tilde{{\textbf{p}}}}$ denotes the vector that each element is the multiplicative inverse of the vector $\textbf{p}$, i.e., ${{\tilde p}_i} = \frac{1}{{{p_i}}}$. The NCF spectrum denotes the transmission efficiency of useful information on the coded bits. In order to quantify the transmission efficiency of useful information, the Euclidean norm of the NCF spectrum can be calculated as
\begin{equation}
 \begin{split}
{\left\| { \gamma(\textbf{x}) } \right\|_2} = \sqrt { \sum\limits_{i = 1}^N {{{ \gamma_i }^2}}}.
\end{split}\label{eq0012}
\end{equation}

Similar with the signal processing, we can regard the rate spectrum as a signal and obtain its energy by
\begin{equation}
\begin{split}
E = \left\| { \gamma(\textbf{x}) } \right\|_2^2 = \sum\limits_{i = 1}^N {{{ {{\gamma_i}}}^2}} .
\end{split}\label{eq0013}
\end{equation}

\section{Optimization of the Rate-profile of PAC Codes}\label{sec4}

In Section III, it was observed that the energy of the NCF spectrum can be adjusted by changing the frozen set. Motivated by this, we design an optimization algorithm to select a good frozen set of PAC codes.

\subsection{Quadratic Constrained Optimization Model}

Since the NCF is a heuristic measure of the transmission efficiency of the useful messages, the energy of the NCF spectrum can be used as the following objective function for the optimization of PAC codes:
\begin{equation}
\begin{split}
q = \mathop {\max }\limits_{{\cal F}} \left\{ E \right\},
\end{split}\label{eq10}
\end{equation}
where $\cal F$ denotes the frozen set and ${\tilde {\textbf{G}}}$ is the compression coefficient matrix. The vector $\textbf{ r(v)}$ is determined by (\ref{eq009}). Therefore, the constraint of the optimization is
\begin{equation}
\begin{split}
\left| \textbf{v} \right| = K, \ \ \ \textbf{v} \in {\mathbb{B}^N},
\end{split}\label{eq0015}
\end{equation}
where $\mathbb{B}$ denotes the binary alphabet. Since the matrix $\tilde{\textbf{G}}\tilde{\textbf{{G}}}^T$ is positive definite, the maximum point of the objective function is on the boundary of the feasible region. With (\ref{eq0015}), the feasible region is determined by the vertices of a polyhedral, which is difficult to determine the upper and lower boundaries. Therefore, a heuristic searching scheme is designed to search for an optimal solution of (\ref{eq10}).

\subsection{Bit-swapping Strategy}

In order to optimize the rate-profile of PAC codes, we propose a bit-swapping strategy to solve the problem given in (\ref{eq10}). The objective function can be regarded as a metric:
\begin{equation}
\begin{split}
\varphi =  r\textbf{(v)}\tilde{\textbf{G}}{\tilde{\textbf{G}}^T}{r\textbf{(v)}^T}.
\end{split}\label{eq022}
\end{equation}

By exchanging the positions of some frozen bits and non-frozen bits, the frozen set with large $\varphi$ will be selected as the optimized frozen set. Since the feasible region contains $\frac{{N!}}{{\left( {N - K} \right)!K!}}$ vertices, it is impractical to search on all the possible candidates. Therefore, we propose to select a part of bits as the search space for the bit-swapping strategy.

According to \cite{PAC}, the RM-design of the PAC codes shows the best performance among three rate-profile methods. Therefore, we adopt the frozen set obtained by the RM-design rate-profile as the starting point of the bit-swapping. In addition, the RM-design sorts the indexes according to the RM score function\cite{PAC}, where the RM score function is
\begin{equation}
\begin{split}
{\cal D}\left( i \right) = \omega \left( {i - 1} \right).
\end{split}
\end{equation}

The $\omega \left( {i - 1} \right)$ denotes the Hamming weight of the binary representation of $i-1$. We adopt the indexes with medium RM scores as the search space. Based on simulation experiments, we find that the swapping among a few bits would improve the performance of PAC codes. Let $M$ denote the maximum number of the swapped bits. Let ${\cal F}_{\text{opt}}$ denote the optimized frozen set. The optimized rate-profile algorithm of $(N, K)$ PAC codes is summarized by:

\begin{enumerate}
  \item \textbf{Initialization}: Calculate the RM scores for all \emph{N} indexes. Sort the \emph{N} indexes according to their RM scores and select $N-K$ indexes with small RM scores as the starting frozen set ${\cal F}_0$. Calculate the metric $\varphi_0$ corresponding to the starting frozen set by (\ref{eq022}).
  \item \textbf{Starting Point}: Select $M$ indexes in the frozen set ${\cal F}_0$ with large RM scores as a new set $\Phi$. And select $M$ indexes with small RM scores in the non-frozen set ${\cal F}^c_0$ as a new set $\Psi$. The $\Phi$ and $\Psi$ constitute the search space. Set $\varphi$ = $\varphi_0$ and ${\cal F}_{\text{opt}}$=${\cal F}_{\text{0}}$.
  \item \textbf{Swapping}: Exchange the elements between $\Phi$ and $\Psi$, then a new frozen set ${\cal F}_{\text{temp}}$ is obtained. Calculate the metric corresponding to the new frozen set ${\cal F}_{\text{temp}}$ by (\ref{eq022}), which is denoted by $\varphi_{\text{temp}}$.
  \item \textbf{Updating Metrics}: If the metric $\varphi_{\text{temp}}>\varphi$, the optimization will be updated by $\varphi$=$\varphi_{\text{temp}}$ and ${\cal F}_{\text{opt}}$=${\cal F}_{\text{temp}}$. Otherwise, the optimized frozen set and the metric will remain unchanged.
  \item \textbf{Looping}: Return to Step 3 and obtain the next candidate by swapping the elements between $\Phi$ and $\Psi$. After all the elements in $\Phi$ and $\Psi$ have been swapped, the optimized frozen set is ${\cal F}_{\text{opt}}$.
\end{enumerate}

\section{Simulation Results}\label{sec5}

In this section, the frame error rate (FER) performance of the proposed optimized PAC codes is evaluated. The (128, 64) PAC codes with generator polynomials $\textbf{g}$ = 0o133, 0o177 are used in simulation. Three cases are considered, where $M=0,2,4$. With $M=0$, the frozen set is constructed by the RM rate-profile. When $M=2$, the search space is $\Phi$=$\{105, 113\}$, $\Psi$=$\{16, 24\}$. With $M=4$, the search space is $\Phi$=$\{99, 101, 105, 113\}$, $\Psi$=$\{16, 24, 28, 30\}$. Table \ref{tab1-1} indicates the $\varphi$ metric of PAC codes with the convolutional generator polynomial 0o177 under different swapping patterns. We find that the swapping ($\{99, 101, 105, 113\}\leftrightarrow \{16, 24, 28, 30\}$) corresponds to the highest metric among the three cases.

\begin{figure}[ht]
\begin{center}\hspace{-0.381cm}
\centering
\includegraphics[scale=0.46389336]{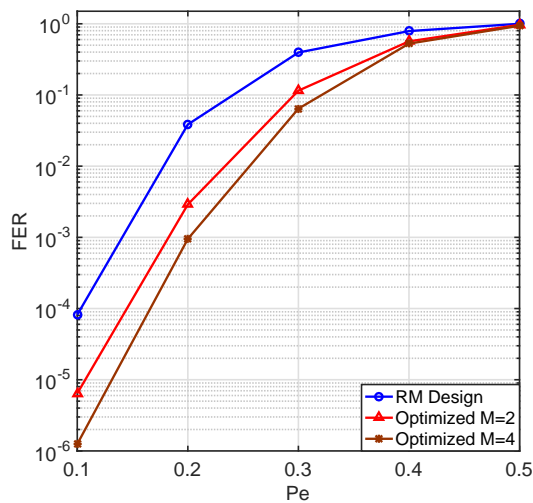}
\caption{FER performance of PAC codes in BEC.}
\label{F003FER128}
\end{center}
\end{figure}

Fig. \ref{F003FER128} shows the FER performance of the PAC codes with the convolutional generator polynomial 0o177 in BEC. The PAC codes are decoded by the successive cancellation decoder. In Fig. \ref{F003FER128}, the PAC codes with $M=4$ have the lowest FER and the highest $\varphi$. When the probability of erasure error is $0.2$, the FER of the PAC codes with the optimized rate-profile is $10^{-3}$, while the FER of the PAC codes with the RM rate-profile is higher than $10^{-2}$. The superiority of the optimized rate-profile benefits from the adequate utilization of the polarized channel capacity, where the bits with medium RM scores play an important role indeed.

\begin{table}[ht]
\renewcommand{\arraystretch}{1.3}
\caption{NCF Metric With Different Rate-Profiles.}
\centering\begin{threeparttable}
\begin{tabular}{|c|c|c|c|}
\hline
M & 0 & 2 & 4 \\
\hline
$\varphi$&$73.81$&$82.21$&$83.61$\\
\hline
\end{tabular}
\end{threeparttable}
\label{tab1-1}
\end{table}

\begin{figure}[H]
\begin{center}\hspace{-0.381cm}
\centering
\includegraphics[scale=0.49336196536]{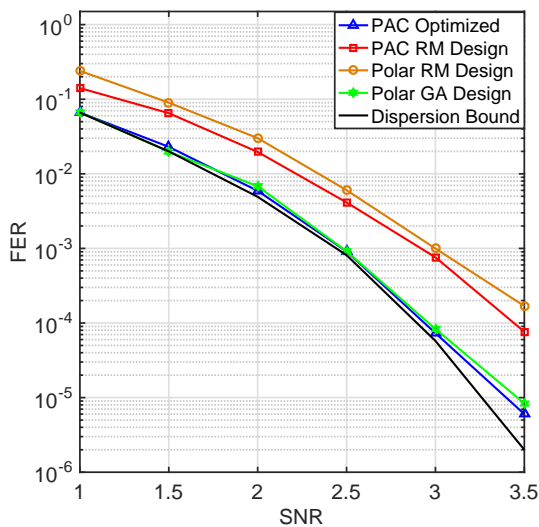}
\caption{FER of PAC and polar codes with list decoding in AWGN channel.}
\label{F008}
\end{center}
\end{figure}

Fig. \ref{F008} depicts the FER performance of the (128, 64) PAC codes and polar codes in the additive white Gaussian noise (AWGN) channel, where the generator polynomial of the convolutional transform is 0o133. The PAC codes and polar codes are concatenated with the cyclic redundant check (CRC) codes, where the generator polynomial of the CRC code is $0$xA$6$. Then they are decoded by the SCL decoder with list size 32. The FER performance of the optimized PAC codes with $M=4$ and the performance of the polar codes based on the Gaussian approximation (GA) construction are shown in Fig. \ref{F008}, where the difference between them is negligible. The GA construction of polar codes is a real time construction and the frozen set needs to be calculated separately under different signal-to-noise ratios (SNR)s. However, the frozen set obtained by the optimized rate-profile remains the same with different SNRs. Moreover, the NCF metric of PAC codes with the RM rate-profile is $\varphi=82.52$. The NCF energy of the optimized PAC codes is $\varphi=87.45$, which is larger than that of the PAC codes with the RM rate-profile. Meanwhile, the PAC codes with the optimized rate-profile show a better decoding performance than the PAC codes with the RM rate-profile. Specifically, at BLER=$10^{-3}$, the performance gain of the optimized PAC codes against the PAC codes with the RM rate-profile is about 0.35 dB.

\section{Conclusion}\label{sec6}

This paper proposes a novel method to analyze and optimize the construction of PAC codes. First, the NCF is proposed to heuristically measure the transmission efficiency of useful messages. Furthermore, we find that the NCF distribution of the PAC codes should match the capacity distribution after finite-length polar transform, which provides a sight into the performance advantages of the PAC codes against polar codes. Then, a quadratic optimization model is established to optimize the PAC codes. A bit-swapping strategy is designed to obtain the optimized rate-profile of PAC codes. Simulation results verify the advantages of the proposed rate-profile optimization. In addition, the NCF also provides a basis for the optimization of the generator polynomials. In the future, we will optimize the generator polynomials of the PAC codes.

\ifCLASSOPTIONcaptionsoff
  \newpage
\fi

\end{document}